\documentstyle[12pt,agums]{article}

\lefthead{ALVAREZ ET AL.}

\righthead{EFFECTS OF SMALL SCALE VARIABILITY ON OCEANIC CURRENTS}

\received{September~01,~1999}
\revised{November~23,~1999}
\accepted{January~19,~2000}

\paperid{}

\cpright{AGU}{1999}

\authoraddr{A. Alvarez,
SACLANT Undersea Research Centre, 19138 San Bartolomeo, La Spezia,
Italy. (e-mail: alvarez@saclantc.nato.int)}

\authoraddr{E. Hern\'andez-Garc\'{\i}a and J. Tintor\'e,
Instituto Mediterr\'{a}neo de Estudios Avanzados, CSIC-Universitat
de les Illes Balears, 07071 Palma Mallorca, Spain.   (e-mail:
emilio@imedea.uib.es; dfsjts0@ps.uib.es)}

\begin{document}

\title{On the effect of small-scale oceanic variability
\\on topography-generated currents}

\author{A. \'Alvarez}

\affil{SACLANT Undersea Research Centre,
 La Spezia, Italy}

\author{E. Hern\'andez-Garc\'{\i}a and J. Tintor\'e}

\affil{Instituto Mediterr\'{a}neo de Estudios Avanzados,
CSIC-Universitat de les Illes Balears, Palma de Mallorca, Spain}

\begin{abstract}

Small-scale oceanic motions, in combination with bottom
topography, induce mean large-scale along-isobaths flows. The
direction of these mean flows is usually found to be anticyclonic
(cyclonic) over bumps (depressions). Here we employ a
quasigeostrophic model to show that the current direction of these
topographically induced large-scale flows can be reversed by the
small-scale variability. This result addresses the existence of a
new bulk effect from the small-scale activity that could have
strong consequences on the circulation of the world's ocean.

\end{abstract}

\begin{article}
\section{Introduction}

Small-scale ocean motions have an important effect on oceanic
flows several orders of magnitude larger than them. The
best-known bulk effect of small-scale processes is a
substantial contribution to the transport of heat, salt, momentum,
and passive tracers in all parts in the world's oceans. This
effect is usually included in ocean circulation models by
modifying the transport and mixing properties of the fluid from
their molecular values to larger ones, giving rise to
eddy-diffusion approaches of increasing sophistication and
predictive power \cite{Ne94}.
The transport processes parametrized by these
effective changes of the diffusive fluid properties have been
shown to control important aspects of the Earth's climate \cite{Da94}.

Beyond eddy diffusion approaches, physical effects of small-scale
activity are still poorly understood. For this reason, the nature
and variability of small-scale oceanic motions have been
exhaustively examined in different oceanographic contexts
\cite{Wu95}. Given the nature of small-scale activity --
disordered, fluctuating and turbulent -- a contribution to
diffusion and dispersion effects is obvious on physical grounds.
But a more coherent influence of processes occurring at
small-scales on large scales motions is unexpected unless some
oceanographic factor is able to get a significant mean component
out of the fluctuating behavior. Bottom topography is one of such
factors breaking the symmetry of the fluctuation statistics, and
thus provides a dynamical link for energy transfer from the small
to the large scale.

Evidence has been accumulated in the last decade showing that mean
flows following the topographic contours are often found in the
vicinity of topographic features of different scales
\cite{Kl89,Po91,Ma94,Br95}.
These topographically
generated currents have been shown to influence both local and
global aspects of the Earth's climate \cite{De98}. For example, large-scale
motions related to topographic anomalies have been found in the
North and South Atlantic playing a major role in determining
regional circulation and climatic characteristics \cite{Lo95,So95}.

Coriolis force, topography and fluctuations have been pointed out
as the main ingredients to generate these along-isobaths coherent
motions \cite{Al97,Al98a,Al98b}. Briefly, Coriolis force links topography to
the
dynamics of ocean vorticity. Thus changes in ocean depth provide a
symmetry breaking factor distinguishing according to their
vorticity content otherwise isotropic mesoscale fluctuations.
The result is that the mean effect of small-scale fluctuations
does not
average to zero yielding the existence of mean
flows. Finally, the topographic structure determines the
circulation patterns of the originated currents.

On the basis of present knowledge, anticyclonic (cyclonic)
tendencies are expected over bumps (depressions) for generated
mean flows over topography. However, \markcite{{\it Alvarez et
al.}} [1998] pointed out that these circulations tendencies could
be strongly dependent on the properties of the small-scale
variability. They theoretically addressed the possibility that the
above mentioned circulation pattern could be even reversed
(cyclonic (anticyclonic) circulations over bumps (depressions)) by
the action of the small scales. The same effect is predicted when
considering bottom friction instead of viscosity as the damping
mechanism \cite{Al99}. The present Letter attempts to elucidate,
by means of computer simulations, if the direction of these mean
flows is sensitive to the statistical characteristics of the
small-scale, as it was argued on theoretical grounds.

\section{Model and results}

To explore in detail the possible relations between large and
small scales in the presence of topography an ideal ocean
represented by a single layer of fluid subjected to
quasigeostrophic dynamics will be considered. Baroclinic effects
which in real oceans give rise to meso-small scale activity are
modeled here by an explicit stochastic forcing with prescribed
statistical properties \cite{Wi78}. This term might also be
considered as representing any high frequency wind forcing
components and other processes below the resolution considered in
the numerical model. This random forcing, in combination with
viscous dissipation, will bring the ocean model to a statistically
steady state. While highly idealized, the simplifying modeling
assumptions above arise from our interest in isolating just the
essential processes by which small-scale variability leads to
topography-generated currents.

Within our approximations, the full ocean dynamics can be described
by \cite{Pe87}:

\begin{equation}
\label{eq}
{\partial \nabla^2 \psi \over \partial t} +
\left[ \psi , \nabla^2 \psi +h \right] =
\nu \nabla^4 \psi + F \ ,
\end{equation}

The ocean dynamics described by Eq.~(\ref{eq}) is an f-plane
quasigeostrophic model where $\psi({\bf x},t)$ is the
streamfunction, $F({\bf x},t)$ is the above mentioned stochastic
vorticity input, $\nu$ is the viscosity parameter and $h=f \Delta
H/H_0$, with $f$ the Coriolis parameter, $H_0$ the mean depth, and
$\Delta H({\bf x})$ the local topographic height over the mean
depth. The Poisson bracket or Jacobian is defined as
\begin{equation}
\label{jac}
[A,B]= {\partial A \over \partial x} {\partial B \over \partial y} -
{\partial B \over \partial x}{\partial A \over \partial y}\ .
\end {equation}

A set of numerical simulations has been carried out to determine
the dependence of the large scale pattern circulation on the
structure and variability of the small-scales. The description of
the numerical model and different parameters employed in the
simulations are summarized in Appendix A. A randomly generated
bottom topography is used in all the cases. As a way of changing
in a continuous manner the statistical properties of  the forcing
$F({\bf x},t)$ we assume it to be a Gaussian stochastic process of
zero mean, white in time, and spatial spectrum given by $S(k)
\propto k^{-y}$, where $k$ is the wavenumber. A positive exponent
$y$ represents relative-vorticity fluctuations more dominant at
the large scales, whereas negative $y$ represents fluctuations
dominant at the smaller scales. The distribution of fluctuation
variance among the scales can thus be controlled by varying $y$.
The spectrum of the energy input corresponding to the above
stochastic vorticity forcing is also white in time, with a
wavenumber dependence given by the relation $E(k)=S(k)~k^{-2}$. We
have started first considering a situation where the small-scale
variability is described by $S(k) \propto k^{0}$. This power-law
has been observed for vorticity forcing induced by winds in the
Pacific ocean \cite{Fre85}. The model has been integrated until a
stationary state is achieved. Figure 1b shows the mean currents
obtained from this specific simulation. In the mean state the
currents do not average to zero, despite the isotropy of the
fluctuations and dissipation. Instead the final mean state is
characterized by the existence of large-scale mean currents
strongly correlated with bottom topography. The spatial
correlation coefficient between the streamfunction and the bottom
topography is for this case $0.85$. This positive spatial
correlation implies the existence of mean anticyclonic (cyclonic)
circulations over bumps (depressions). As a next step, we have
modeled the action of small-scales as a noisy process with a
correlation described by the power-law $k^{4.8}$. This spectrum
describes a situation where the small-scale variability is more
energetic than the one induced by the previous $k^{0}$ power-law.
The response of the system is drastically changed by this
small-scale activity. As shown in Figure 1c the mean state of the
ocean displays a pattern of circulation practically uncorrelated
with bottom topography. Specifically, the spatial correlation
coefficient is $0.091$. Increasing again the exponent of the power
law to $S(k) \propto k^{6}$, we obtain the generation of mean
currents anticorrelated with bottom topography, as it can be
observed from Figure 1d. The spatial correlation coefficient is
$-0.77$ in this case of high small-scale activity, indicating the
existence of mean cyclonic (anticyclonic) motions over bumps
(depressions). Note that Figures 1c and d display
topography-generated currents much weaker than those obtained for
the $k^{0}$ power-law case, Figure 1b. This feature comes from the
scale-selective character of the viscosity, more efficient at
small scales where the forcing energy is most localized in the
$k^{4.8}$ and $k^{6}$ cases. Besides this effect, it should also
be mentioned that forcing with a $k^{0}$ spectral power-law
directly provides more energy input to the large-scale components
than the $k^{4.8}$ and $k^{6}$ forcings. Additional numerical
simulations, for different initial conditions and noise and
topography realizations, consistently confirm the results of the
simulations presented in Fig.~1, that is the sensibility of the
large-scale circulations not only to the particular structure of
the underlying topography but also to the characteristics of the
small-scale variability of the environment. In particular, as the
small-scale content in the vorticity forcing is augmented with
respect to the large-scale one, mean currents are always seen to
reverse direction.

\section{Conclusion}

On the basis of the property of potential-vorticity conservation,
anticyclonic (cyclonic) motions are traditionally expected over
topographic bumps (depressions) \cite{Pe87}. If potential
vorticity is not preserved because of the presence of some kind of
forcing mechanism, then different circulation patterns can be
generated. Small-scale activity constitutes a systematic and
persistent forcing of the circulation in the whole ocean. Due to
the relatively small and fast space and time scales that
characterize this variability, the physical characteristics of
this forcing are usually described in terms of their statistical
properties \cite{Wi78}. In other words, small-scale activity can
be considered as a fluctuating background in which the large-scale
motions are embedded. The relevance of the role played by this
fluctuating environment in modifying the transport and viscous
properties of the large scales is widely recognized. Beyond these
diffusive and viscous effect of the small-scale activity, the
numerical results presented in this Letter show that in the
presence of bottom topography, statistical details of the
variability of the small-scales can induce different large-scale
oceanic circulations. The strength of this effect will be affected
by the spectral characteristics of the topographic and forcing
fields as well as by the real baroclinic nature of the ocean.
Preliminary computer simulations indicate that the strength of the
mean currents increases when the topography contains more
proportion of large-scale features. This effect and the dependence
of the current direction on the forcing spectral exponent,
presented in Fig. 1, nicely validates the theoretical results in
\cite{Al98b}. Extension of the theoretical methods to the
baroclinic case is in progress. However, a complete analysis of
the influence of different forcings and topography shapes can only
be addressed numerically. The shape of the topography was already
shown to play a fundamental role in the energy transfer between
different scales in baroclinic quasigeostrophic turbulence
\cite{Tre88}. Finally, the results shown in this Letter stress the
need for a better observational characterization of the space and
time variability of oceans at small-scales in order to achieve a
complete understanding of the large-scale ocean circulation.

\appendix

\section{Appendix A: Numerical model description}

Numerical simulations of Eq.~(\ref{eq}) have been conducted in a
parameter regime of geophysical interest. A value of $f=10^{-4}
s^{-1}$ was chosen as appropriate for the Coriolis effect at mean
latitudes on Earth and $\nu =200  m^2 s^{-1}$ for the viscosity, a
value usual for the eddy viscosity in ocean models. We use the
numerical scheme developed in \markcite{{\it Cummins}} [1992] on a
grid of $64 \times 64$ points. The distance between grid points
corresponds to 20 km, so that the total system size is $L=1280$
km. The algorithm, based on Arakawa finite differences and the
leap-frog algorithm, keeps the value of energy and enstrophy
constant when it is run in the inviscid and unforced case. The
consistent way of introducing the stochastic term into the
leap-frog scheme can be found in \markcite{{\it Alvarez et al.}}
[1997]. The amplitude of the forcing
%,$D=1\times 10^{-7}\  m^2/s^2$
has  been chosen in
order to obtain final velocities of several centimeters per second.
The topographic
field is randomly generated from a isotropic spectrum containing, with
equal amplitude and random phases, all the Fourier modes corresponding to
scales between 80 km and 300 km.
The model was run for
$5 \times 10^5$ time steps (corresponding to 206 years) after a
statistically
stationary state was reached.  The streamfunction
is then averaged during this last interval of time.

\acknowledgments

Financial support from CICYT (AMB95-0901-C02-01-CP and
MAR98-0840), and from the MAST program MATTER MAS3-CT96-0051 (EC)
is greatly acknowledged. Comments of two anonymous referees
are also greatly appreciated.

{}
\end{article}

\newpage

\begin{figure}
        \caption{a) Random bottom topography. Maximum
and minimum topography heights are $500 m$ and $-599 m$,
respectively, over an average depth of $5000 m$. b) Computed mean
streamfunction $\psi({\bf x},t)$ in $m^{2} s^{-1}$ for the case
when the small-scale variability is described by a spectral power
law $k^0$. Bottom topography levels have been superimposed (black
lines) as reference over the streamfunction field. The strong
correlations between the streamfunction and topography are clear
from this figure. c) Same as b) but with power spectra law
$k^{4.8}$. In this case the flow remains practically uncorrelated
with the underlying topography (black lines). d) For $k^6$ the
flow is anticorrelated with the topography. }
\end{figure}

\end{document}